Influence of muscle preactivation of the lower limb on impact dynamics in case of frontal collision.

Martine Pithioux[1], Pascale Chavet[1], Nancy St-Onge [2], Caroline Nicol[3]

[1] Laboratory of Aerodynamics and Biomechanics of Motion (LABM), USR 2164 CNRS-Université de la Méditerranée, CP 918, 163 av. de Luminy, Marseille 13288 cedex 09, France.

[2] Jewish Rehabilitation Hospital, 3205, Place Alton Goldbloom, Chomedey, Laval, Québec, H7V 1R2, Canada.
School of Physical and Occupational Therapy, McGill University, 3654, Promenade Sir-William-Osler, Montreal, Quebec H3G 1Y5, Canada.

[3] Physiological Determinants of Physical Activities (DPAP), UPRES EA 3285 Université de la Méditerranée, CP 910, 163 av. de Luminy Marseille 13288 cedex 09, France.

**Corresponding authors :**

**Martine Pithioux or Pascale Chavet:** Laboratory of Aerodynamics and Biomechanics of Motion (LABM), USR 2164 CNRS-Université de la Méditerranée, CP 918, 163 av. de Luminy, Marseille 13288 cedex 09, France.

*Email : pithioux@morille.univ-mrs.fr    chavet@morille.univ-mrs.fr*

Phone : +33 491 26 60 30    Fax : +33 491 41 16 91

**Abstract**

Accidentology or shock biomechanics are research domains mainly devoted to the development of safety conditions for the users of various transport modes in case of an accident. The objective of this study was to improve the knowledge of the biomechanical behaviour of the lower limb facing sudden dynamic loading during a frontal collision. We aimed at establishing the relationship between the level of muscular activity prior to impact, called 'preactivation', of the lower limb extensors and the mechanical characteristics of impact. Relationships were described between the level of preactivation, the impact peak force values, the minimum force after unloading and the associated loading and unloading rates. The existence of reflex mechanisms that were affected by the level of voluntary muscular preactivation for the lower limb muscles was demonstrated. In conclusion, the existence of specific mechanism acting mainly at the knee level may result from the level of preactivation. Muscle behavior has to be included in numerical models of the human driver to better evaluate the overall stiffness of the body before and at impact.

1. INTRODUCTION

The increasing use of passive restraint systems such as seatbelts and airbags for drivers in car occupants leads to an apparent increase of lower limb injuries resulting from automotive accident [1] and might become more prevalent in years to come [2, 3]. Biomechanical studies of collision tests contributed to improve the knowledge of the effects of these systems on internal and external integrity of human body. Injuries inferred by the collision were most of the time indirectly studied via post mortem human surrogates (PMHS) or dummies paradigms in simulated or real collisions [4, 5]. However, recent epidemiology studies [6] pointed out the important occurrence of lower limb injuries which are not predicted by the simulation models. None of the existing experimental paradigms considered the effect of muscle activation especially the functional responses of the musculo-skeletal system. Bracing occurs when a car occupant is aware of an imminent collision and should affect musculoskeletal responses as the mechanical characteristics of impacts. Muscle activation and more particularly muscle stiffness has often been considered as a regulated property of the neuromuscular system. From a mechanical perspective, stiffness describes the ratio of force response of a material to an imposed change in that resists mechanical stretch perturbation. This resistance limits joint motion, minimizing ligamentous strain and subsequent injury. Thus, control of lower limb muscle stiffness may be a primary mechanism by which potential injury resulting from collision could be minimized. Specifically, as it was shown in gait biomechanics studies, the muscular preactivation appears to be a preparatory requirement both for the enhancement of force production during the braking phase and for timing of muscular action with respect to ground contact [7]

The long-term objective of such study is to supply numerical models of car occupant with muscular response behavioral law at impact based on biomechanical and neuromuscular experimental data. Volunteers will be preferred to PMHS or dummies as it was the case in the literature. The interest of such study is to understand the effect of lower limb muscle preactivation on impact dynamics.

Even if it is obvious that an automotive collision exceeds the functional human boundaries [1, 8] the imminent occurrence of a collision may generate protective behaviours both conscious and reflex [8, 9]. Similar adaptations are reported for lower velocity range obtained from human sport performance, not included the motorsports [10-17]. These studies used a specific ergometer, the sledge [10], and emphasized the importance of muscle preactivation on impact force [18]. However, these observations were conducted in rebound performance testing with impact forces largely inferior (impact velocity from 1 to 4 $m.s^{-1}$) to those generated by an automotive collision (impact velocity from 8 to 14 $m.s^{-1}$). Most of the studies used fatigue protocols [11-19] based on cycling exercises or using sledge ergometers to induce changes in the preactivation level prior to impact. These studies showed the occurrence of protective reflex mechanisms, even in case of sub-maximal impacts. In a frontal car crash situation, protective mechanisms are expected to occur as the occupant is aware of the coming up collision.

In the current study, sub-maximal impact velocity was used to test volunteers instead of functionally reductive PMHS or dummies. The inclusion in the existing numerical model (HUMOS1, European Project (Reference: BRPR970475)) of a behavioral law, characterising the effect of muscle action to be closer to reality, will allow to take into account muscular activation modulation around impact which will be expressed as a variable resistance force output. The activity in 4 limb muscles was monitored and the volunteers were asked to maintain a specific percentage of the global muscle activity before impact.

The use of the sledge allowed the analysis of collision effects in the seating position. It also presents the interest of an adjustable antero-posterior inclination, which allows the control of the sledge seat velocity until impact. Other variables may be adjusted and controlled before impact: the initial position of the volunteer-seat, the angular positions and the stiffness of the involved body segments.

In the present study a guiding device was affixed to the sledge seat to allow the subjects to push with their right foot to generate a predetermined level of force associated with muscle preactivation before impact: 25, 50, and 75% of the maximum voluntary force. The participant was raised by the experimenter to a supra maximal dropping height corresponding to 200% of his individual maximal rebound height. The resulting impact velocity was on the average 4 $m.s^{-1}$. Although lower than the accidentology impact threshold, this velocity limitation was necessary to protect the volunteer from potential musculoskeletal injuries. Based on the rebound literature such testing conditions are relevant to analyse the effect of muscle preactivation in case of a forthcoming frontal collision.

## 2. MATERIALS AND METHOD

Ten male adults (mean age 27) whose mass and height fitted with the European $50^{th}$ male percentile (1.78 m, 77 kg) volunteered for this study. At the time of the experimentation they were free of pain and injuries to their lower extremities. The experimental protocol was approved by the Ethics Committee (n° 04008- CCPPRB Marseilles 2).

### 2.1. EXPERIMENTAL SET-UP

Impacts were delivered to the volunteers using a specific sledge ergometer [19] in the sitting position. A lower limb guiding device developed by our team was affixed to the seat of the sledge and allowed the study of the effect of preactivation level on musculoskeletal behavior as well as on mechanical characteristics of impact.

### 2.1.1. Sledge ergometer

The *sledge* ergometer includes a car seat on rails (Figure 1). The inclination of the rails can be adjusted to control for velocity at impact. There is a force plate (University of Jyväskylä, 585*430*55 mm) at the bottom of the sledge, orthogonal to the rails and a potentiometer (Leine & Linde) affixed to the seat to measure displacement. In the present study, the angle between the incline plane and the horizontal floor was set at 26° to obtain an impact velocity of approximately 4 $m.s^{-1}$. The participant was secured onto the seat by a race car harness to avoid flexion of the trunk. The perpendicular reaction force at the impact was recorded by the force plate.

### 2.1.2. The guiding device

This device (Figure 2) allowed subjects to preactivate leg muscles to predetermined levels by pushing on a foot support with respect to the functional integrity of the lower limb. In addition it permitted to maintain the limb joints in determined angular positions prior to impact.

This device consisted of a metallic structure aligned with the seat base. The aluminium plate at the bottom of the guiding device was designed for foot support and was equipped with a strain gauge sensor (Wheatstone bridge). The volunteer's right forefoot (from toes to metatarsals) was in contact with the aluminium support. Using this system the three lower limb joints were set at specific angles, and the foot kept parallel to the force plate at impact. In addition, its sliding structure allowed flexion of the leg joints at impact.

The strain gauge signal of the guiding device was monitored by a Virtual Instrument developed with acquisition software *(LABVIEW v.6.0)* and was displayed as a feedback control for the participant on a computer monitor using a numerical slider.

### 2.1.3. Surface Electromyography (EMG)

Disposable bipolar EMG surface electrodes were positioned at a constant inter-electrode distance of 20 mm on selected extensor muscles of the right lower limb: *Vastus medialis (VM), Gastrocnemius medialis (GAM), and Soleus (SOL) muscles* (Figure 3). Skin preparation and positioning of the electrodes were done in accordance with SENIAM recommendations [20].

EMG signals, vertical impact force, force signal from guiding device and seat displacement were synchronised and recorded using a portable system (MEGA 3000E, MEGAElectronics) with a sampling frequency of 2000 Hz, a signal amplification of 412 (EMG preamplifier gain: 375), and an 8-500 Hz band-pass filter (3 dB points).

## 2.2. EXPERIMENTAL PROCEDURES

<u>Maximal Height:</u> Maximal height was found using a squat-jump maximum (Sjmax) test. This test consists in a maximal unilateral jump performed in a sitting position (hip angle about 120°, knee angle set at 90°) on the sledge. These tests were realised without the guiding device. To obtain the individual maximal jump height, the seated participant pushed himself away from the force plate with his right leg. This test was repeated three times. The highest performance was considered to be the maximal height.

<u>- Maximal force and maximal voluntary contractions:</u> The participant was tightly secured on the seat. He was asked to push quickly and as hard as possible against the support of the guiding device to measure maximal force production. Isometric maximal voluntary contractions (MVC) were also recorded for each monitored muscle to record a maximal muscular force output. In the present study the hip, knee and ankle joint angles were kept constant. Three 3-second trials were performed for each muscle as well as for maximal force production. The participant rested for 2 minutes between each contraction.

- Crash test (CT): The dropping height was set at 200% of the maximal height reached in the SJmax test (Figure 4). The participant was secured onto the car seat and had his right foot resting on the guiding device foot support. Four conditions were examined. In the first one, so-called 'free condition', the unique instruction was to actively resist the impact while contacting the force plate. In the three other conditions, a visual feedback allowed the subject to generate either a 25, 50 or 75%, of his maximal lower limb force on the forefoot support of the guiding device (MVC) from the release until the impact. These conditions will be referred to as 25, 50, and 75% preactivation force. Each crash condition was repeated three times. The free condition was performed first, and the three other conditions were performed in random order.

## 3. DATA ANALYSIS

Impact force was normalised relative to the subject's mass. Peak force at impact (C), minimum force after unloading (D), loading rate (A), and unloading rate (B) were measured for each individual experiment from the force plate data (Figure 5).

Displacement of the seat was recorded from the potentiometer and its first derivative was computed.

The EMG data were normalised relative to MVC and band-pass filtered using a zero-lag $4^{th}$ order Butterworth filter (10-350Hz). It was then full-wave rectified and low-pass filtered using a zero-lag 4th order Butterworth with a 75Hz cut off frequency to determine the envelope (SENIAM recommendations [20]). The muscular reflex response to a mechanical stimulus can be characterised by a short latency component called M1 whose occurrence depends on the distance of the monitored muscle to the spinal cord. The reflex latencies for the Vastus Medialis were reported at 30-60 ms [20,21] and at 40-70 ms for the Soleus and Gastrocnemius medialis. Three temporal phases were identified: preactivation (100 ms before impact), central "0-M1" (impact until beginning of reflex response: 0-30ms for VM, 0-40 ms for SOL and GAM) and reflex "M1" phase [20, 22]. The post impact 0-M1 and M1 phases are useful to express the central and reflex drive. For each of the temporal phases, area under the EMG envelop was computed for each individual trial. Results for force generated before impact, impact force, and muscle activation were averaged for the three trials of each condition in each subject.

## 4. RESULTS

### 4.1. Force before impact

The force from the guiding device reflected the overall lower limb force output before contacting the force plate.

The computed values for the three preactivation conditions ranged from 2.9 N/kg to 18.6 N/kg at release (Table 2). The average normalised force at release was 4.8 N/kg, 9.7 N/kg, and 14.5 N/kg for the 25%, 50% and 75% preactivation levels, respectively. However, the mean normalised force generated between release and impact ranked from 4.8 N/kg to 5.4 N/kg. The force output varied with the approach of the collision, the 25% preactivation condition was in average maintained from the release until the collision (25% at release, 24% for mean force). This was not the case nor for 50% condition (50% at release, 35% for mean force) neither for 75% condition (75% at release, 38% for mean force).

### 4.2. IMPACT FORCE

The analysis pointed out the influence of muscle preactivation on the dynamics of impact. The average force associated with their corresponding range was calculated for each level of preactivation (Figure 6). As expected, a relationship was found between the initial force level and the impact peak value (Table 2; Figures 7 and 8). The free testing condition appeared clearly as generating the lowest peak amplitude.

The increase in the level of preactivation matched with the highest magnitude of peaks force. In all experimental conditions, the amplitudes of the normalized peak force values varied from 38 N/kg to 55 N/kg and the time to peak was about 3 ms. It is interesting to observe that the increase of impact force is proportional with the increase of preactivation.

Considering the loading and unloading rates, the analysis pointed out in all experimental conditions that negligible force data scattering was encountered at the first rate of loading. The rate of loading (A) increased with the augmentation of the level of preactivation force (Table 3). The loading rate expressed the shock acceptance by the lower limb's structures or 'passive response' and by the active capacity to provide specific responses, i.e. force increase or plateau. On the contrary, after impact, the rate of unloading (B) declined with the increase of the level of preactivation force (Table 3). In addition, the increase of the level of preactivation increases the magnitude of the force minimum (Figure 9).

### 4.3 MUSCULAR ACTIVATION

In parallel to the analysis of force generated before and after the impact, the activity of three lower-limb muscles was studied. The short latency reflex (M1) was severely affected by the increase of peak loading, as expected. Because the latency of M1 for both GM and SOL was identical, the processed EMG was added and noted Triceps surae (TS), which is functionally relevant. M1 only occurred with 30 ms latency for VM (40 ms for TS (GM+SOL)) from the impact and lasted 30 ms for both VM and TS.

The EMG content observed before M1 is called 0-M1 and mainly represents central drive while M1 should reflect reflex. Despite the equivalent areas for 0-M1 for the lower limb extensors muscles with the preactivation (Figure 10), an EMG burst attributed to a reflex loop facilitation was observed. Moreover, the increase of the level of preactivation corresponded with higher magnitude of the reflex facilitation (Figure 11).

## 5. DISCUSSION

The current study aimed at measuring the effects of preactivation level on impact characteristics. The subjects felt from twice their maximal jumping height to obtain a high impact velocity without causing injury. In the literature, only maximal jumping height of 130% was found [20]. In addition, the overall force generated by the lower limb muscles and measured on the guiding device was chosen to express the percentage of preactivation given to the volunteer as feedback.

### 5.1. EFFECT OF PREACTIVATION ON IMPACT FORCE

One of the objectives of this study was to demonstrate the influence of preactivation on impact peak force. The results validated the existence of such relationship between the level of preactivation and the impact peak force amplitude. The calculated relationships between normalised force output and level of preactivation were not proportional. The averaged increase between testing conditions was 2 N/kg which was considered meaningful in the domain of biomechanics [24]. Actually, the statement of a 'biomechanically' meaningful threshold means that the force difference is perceived by the different proprioception sensors

(muscle spindles, Golgi organs, and cutaneous mechanoreceptors) and by consequence potentially regulated. In other words such increase in peak force value may have a mechanical effect on musculo-skeletal and osteo articular structures.

Otherwise the participant could not or would not maintain the same amount of force on the foot support until the impact. For example 75% preactivation actually corresponded to 38% of maximal force at impact and 50% pre-impact force corresponded to 35% of maximal force. Therefore, the demands of the task (≥50% preactivation) in terms of level of muscular contraction cannot be maintain through the gliding.

Impact force output can be characterized by rates of loading and unloading around the impact peak force. In this study, loading rate before the impact peak was not affected by the level of preactivation which reflected the consistency of the behaviour of passive structures. The unloading rate following the impact peak showed a different temporal pattern with a decline of the value of the slope with the increase of preactivation. This had to be interpreted jointly with the variation of the force minimum which was found to decrease with a decline in preactivation. In fact the volunteer provided a better push off with his impacting leg when the preactivation constraint before impact increased due to the inability to maintain the guiding device. From these results it could be concluded that the lower limb was stiffer when the preactivation was higher at impact due to the increase of muscle activity. On the contrary, after impact, the increase of the level of force preactivation corresponded with the less stiff slope (analysis of the first rate of unloading). Thus the volunteer pushed off with his leg on the force plate to resist from the shock. The minimum force revealed that the increase of the level of preactivation resulted in a greater push off.

## 5.2. Effect of preactivation on EMG

The typical neuromuscular mechanism acting in different types of jumps leads to a dramatic inhibition of M1 reflex for low impact load [18].

The effect of preactivation on EMG was studied in the literature through the Stretch Shortening Cycle principle or behavior demonstrated for cyclic activity such as running and jumping [22, 26 and 27]. The main idea is that increasing the EMG activity before landing will prepare the system in high tuning the myotatic reflex loop whose sensors (the muscle spindles) are sensible to muscle length variation and in case of sudden lengthening will provide EMG bursts to prevent yielding.

Muscle preactivation increased the amount of reflex activity in VM and TS. Compared to the 25% condition VM reflex activity presented an increase of 16% and 26% for the 50% and 75% conditions, respectively. For TS an increase of 23% and 24% for the 50% and 75% conditions, were respectively noted.

The notification of an EMG burst, attributed to reflex loop facilitation providing a specific control of the muscle stiffness, may be partly achieved by segmental reflex potentiation.

As a consequence the higher the level of preactivation was the higher the volunteer's ability to push off from the force plate instead of crashing down. In addition from these results, it could be noted that the most solicited joint is the knee joint.

## 6. CONCLUSION

As expected, the existence of relationships between preactivation and impact characteristics such as impact peak force value, minimum force after unloading, post impact EMG activity was demonstrated.

In definitive the increase of the amplitude of the impact peak force values with a higher loading rate, the increase of reflex facilitation, the critical importance of the knee controlled by the Vastus medialis muscle, was illustrated.

Present results indicate that in the case of a forthcoming collision, the general behavior is oriented towards the enhancement of the overall stiffness of the lower limb when the level of muscular preactivation was voluntarily increased [22].

The mean force generated by the volunteer before impact and after release was different from the force at release; this suggested that he was not able to maintain the level of required force until impact, which might be considered as a protective mechanism aimed at decrease stiffness of the lower-limb. This modulation with the increase of preactivation was provided by an active mechanism i.e. a higher EMG activity of the Vastus medialis to counteract the consequence of the impact passive characteristics. It could be stated that the resulting impact force cushioning is reflected by a lower unloading rate and by a knee flexion following impact. It is worth noting that such EMG increase occurred in the M1 period which was reflex controlled.

Results support the point that it is mandatory to include muscle behavior in numerical models of car occupants to better evaluate the overall stiffness of the body before and at impact. The overall stiffness, discussed in the literature, was in the best case taken into account under the form of a numerical constant whereas its modulation is known to affect largely the mechanical properties of the shock [10-15, 28]. The improvement of numerical models biofidelity is deeply linked with the implementation of a behavioral law describing muscle activity. In conclusion, the understanding of muscle influence on impact loading will help the development, via biorealistic numerical models, of passive and especially active security devices in the automotive domain.

Acknowledgements

This research was supported by the European community in the frame of the GROW project 'Human Models for Safety II'. We also would like to express our grateful thanks to Pr. C.

Brunet for the medical supervision. We also would like to thank S. Seguinel and E. Rabichong for her help in grammar and working corrections.

List of captions

Figure 1: Sledge ergometer

Figure 2: Lower limb guiding device.

Figure 3: Schematic representation of monitored muscles. From left to right : Vastus medialis (VM), Gastrocnemius medialis (GAM) and Soleus (SOL)

Figure 4: Position at release for CT (a), CT at impact (b).

Figure 5: Identification of measured parameters that characterise the impact. The Impact Peak forces value (C) minimum force after unloading (D) rate of loading (A), and rate of unloading (B) are represented. . Time unit= $5.10^{-4}$ s.

Figure 6: Force data for the free condition (a), 25% condition (b), 50% condition (c), 75% condition (d). Mean of all subjects. Averaged force (bold), maximal force (clear dotted) and minimal force (dark dotted) are expressed in N/kg.

Figure 7: Mean force data of all subjects for the 3 conditions (25% in clear, 50% in grey and 75% in black).

Figure 8: Normalized impact peak forces values (N/kg) for free, 25%, 50%, 75% conditions for all participants.

Figure 9: Normalized minimum force after unloading (N/kg) for all conditions.

Figure 10: Identification of 0- M1 and M1 phases of EMG signals and reaction force for the different conditions. EMG is expressed in % of MVC for TS and for VM.

Figure 11: Mean of 0-M1 and M1 phases of EMG signals at 25%, 50%, 75% condition. EMG (%MVC) of VM and TS.

Table 1: Guiding device force for the 25%, 50%, 75% of preactivation. Force output from the sensor of the foot support which is measured at the beginning of the release (Force at release).

Normalised mean force was computed from the time of release until the contact with the force plate.

Table 2: Normalised mean impact peak forces values (N/kg) for free, 25%, 50%, 75% conditions for all participants.

Table 3: Mean loading and unloading rates for 25%, 50%, 75% conditions for all participants.

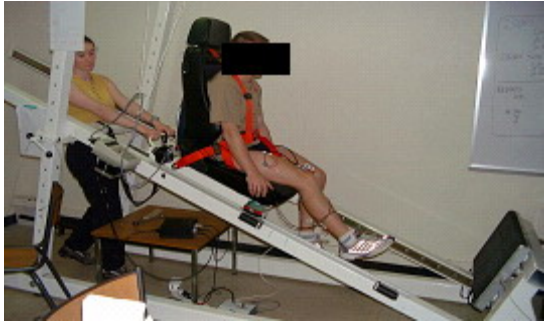

Figure 1.

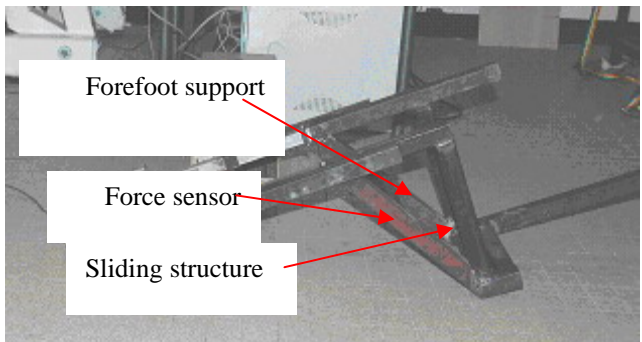

Forefoot support

Force sensor

Sliding structure

Axis bolded to the front and to right side of the seat

Figure 2.

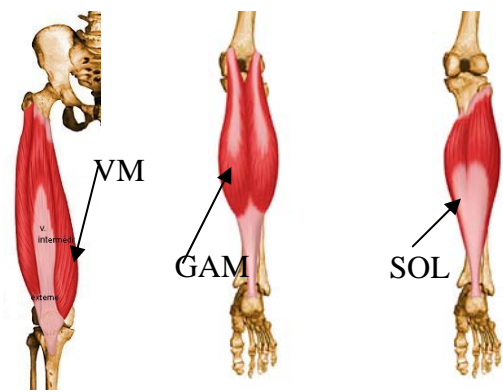

VM

GAM

SOL

Figure 3.

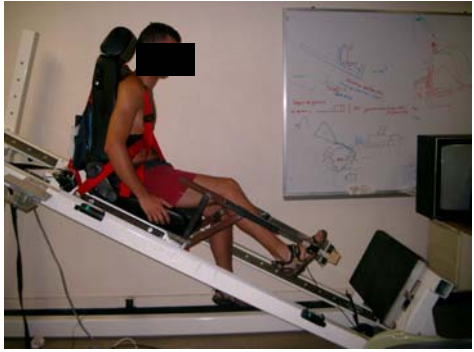 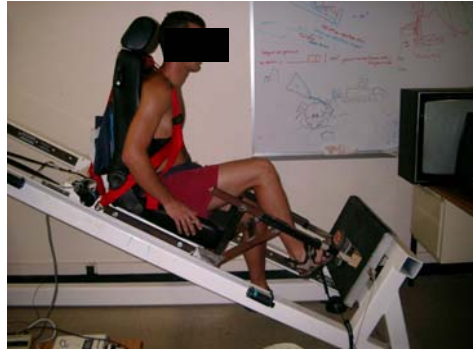

(a) (b)

Figure 4.

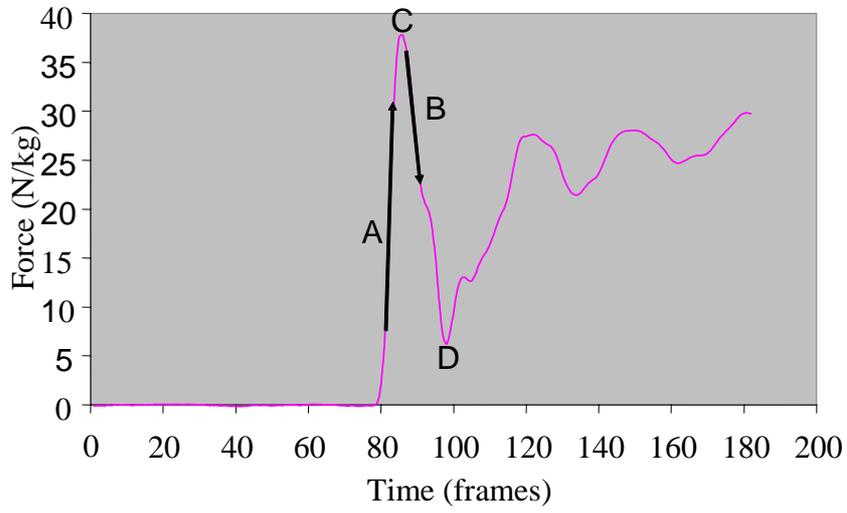

Figure 5.

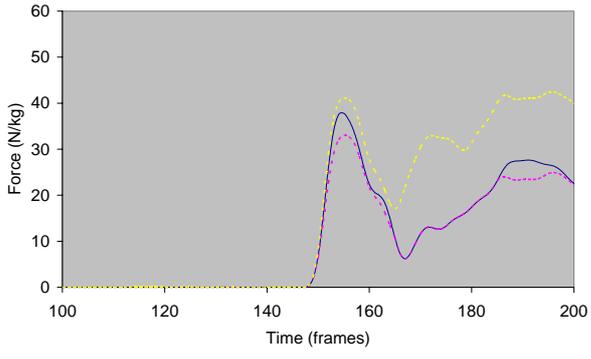

(a)

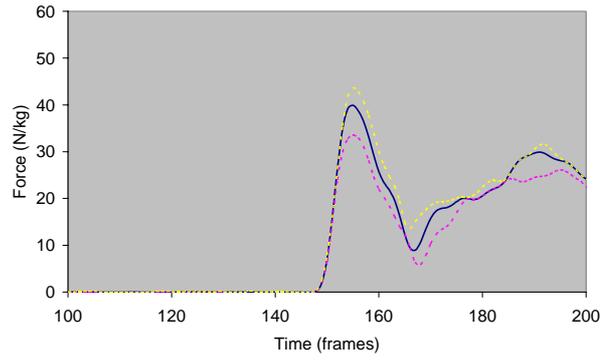

(b)

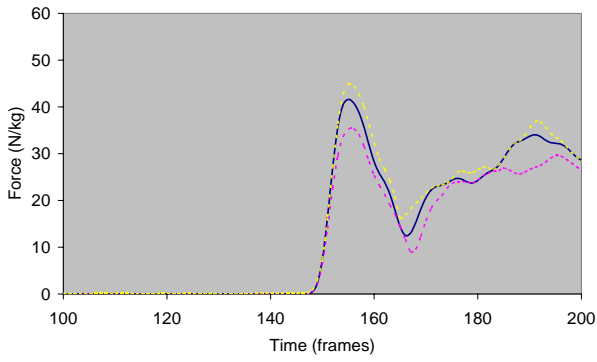

(c)

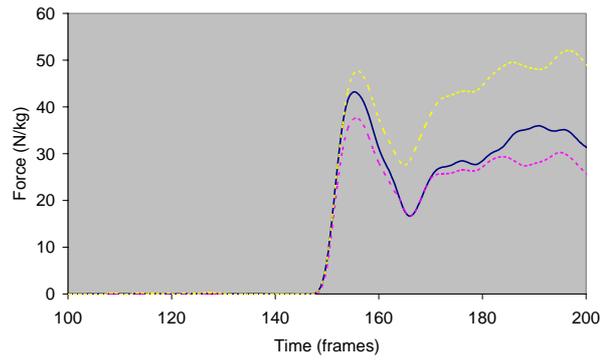

(d)

Figure 6.

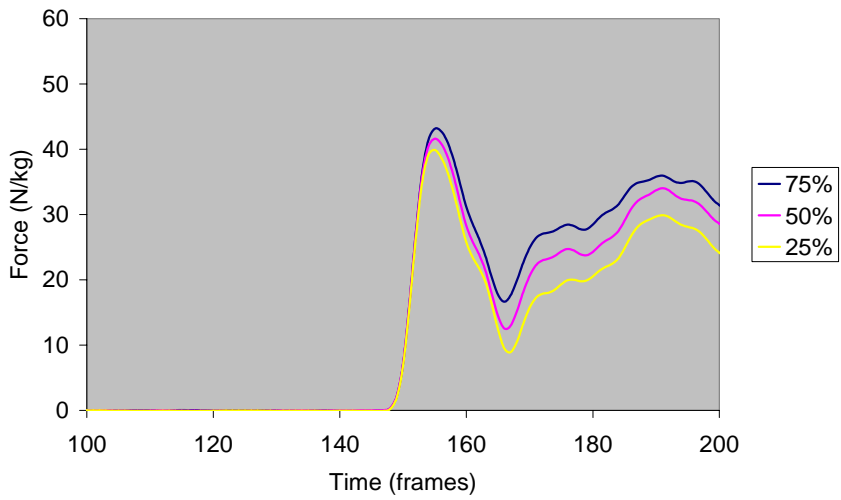

Figure 7.

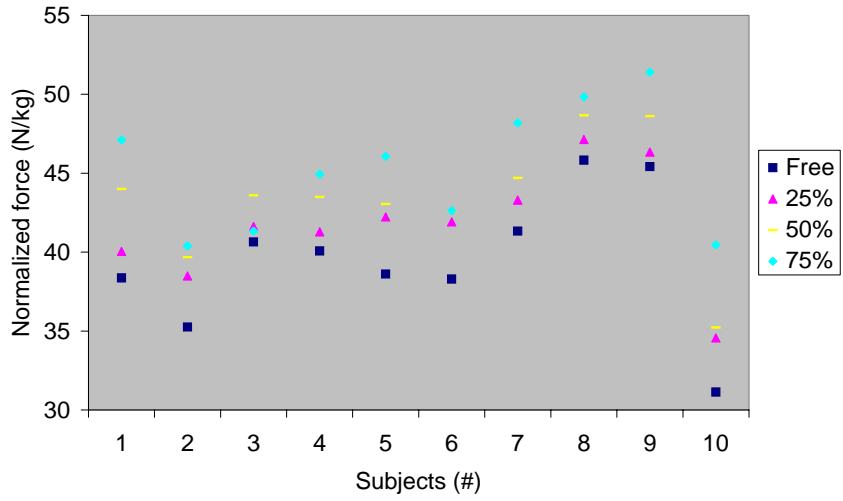

Figure 8.

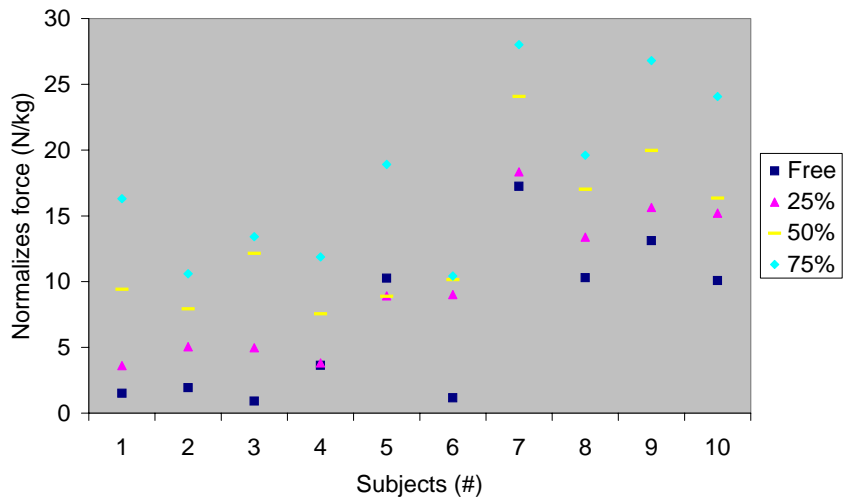

Figure 9.

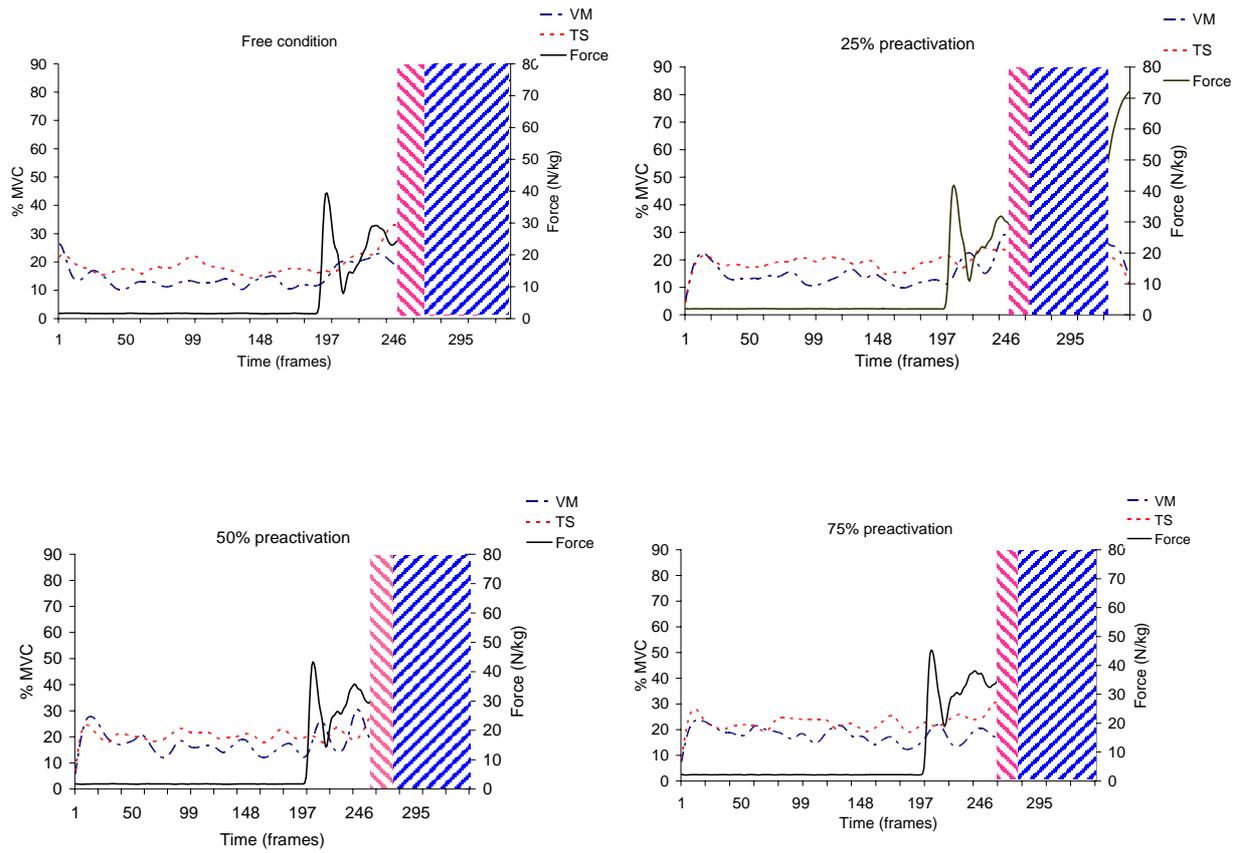

Figure 10.

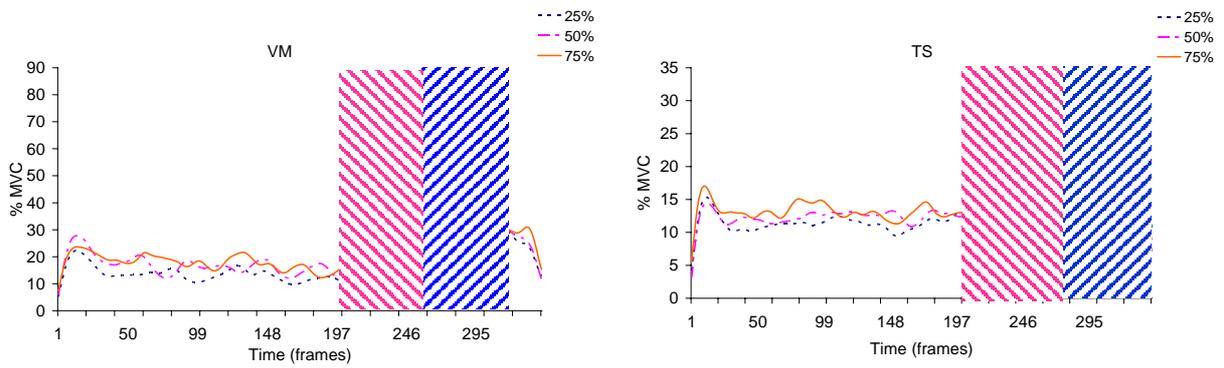

Figure 11.

| Subjects | | 1 | 2 | 3 | 4 | 5 | 6 | 7 | 8 | 9 | 10 |
|---|---|---|---|---|---|---|---|---|---|---|---|
| **Normalized Force at release** | **25% (N/kg)** | 5.8 | 4.5 | 5 | 4.6 | 4.7 | 6.2 | 5.3 | 4.1 | 5.2 | 2.9 |
| | **50% (N/kg)** | 11.6 | 9 | 10 | 9.2 | 9.4 | 12.4 | 10.6 | 8.2 | 10.3 | 5.9 |
| | **75% (N/kg)** | 17.4 | 13.6 | 15 | 13.8 | 14.1 | 18.6 | 16 | 12.3 | 15.5 | 8.8 |
| **Normalized Mean force** | **25% (N/kg)** | 5.7 | 4.5 | 5 | 4.5 | 4.7 | 6.2 | 5.3 | 4 | 5.1 | 2.9 |
| | **50% (N/kg)** | 6.4 | 5 | 5.5 | 5.1 | 5.2 | 6.8 | 5.9 | 4.5 | 5.7 | 3.2 |
| | **75% (N/kg)** | 6.5 | 5.1 | 5.6 | 5.2 | 5.3 | 7 | 6 | 4.6 | 5.8 | 3.3 |

Table 1

| Conditions | Mean normalized impact peak forces (N/kg) |
|---|---|
| **Free** | 44 |
| **25%** | 46 |
| **50%** | 48 |
| **75%** | 50 |

Table 2

| Conditions | **Normalized loading rate at impact (N/kg/s)** | **Normalized unloading rate at impact (N/kg/s)** |
|---|---|---|
| 25% of preactivation | 1410 ± 186 | -484 ± 67 |
| 50% of preactivation | 1413 ± 193 | -477 ± 58 |
| 75% of preactivation | 1429 ± 253 | -454 ± 61 |

Table 3